\documentstyle[pra,aps,epsf,twocolumn,tighten]{revtex}

\relax
\begin{document}

\title{Realization of a 5-Bit NMR Quantum Computer\\
Using a New Molecular Architecture}

\author{R. Marx}

\address{Institut f\"ur Organische Chemie,
J. W. Goethe-Universit\"at, Marie-Curie-Str.\ 11,\\
D-60439 Frankfurt, Germany}

\author{A.~F. Fahmy}

\address{Biological Chemistry and Molecular Pharmacology,
Harvard Medical School,\\ 240 Longwood Avenue, Boston, MA 02115, USA}

\author{John M. Myers}

\address{Gordon McKay Laboratory, Division of Engineering and Applied
Sciences,\\ Harvard University, Cambridge, MA 02138, USA}

\author{W. Bermel}

\address{Bruker Analytik GmbH, Silberstreifen, D-76287
Rheinstetten, Germany}

\author{S.~J. Glaser}

\address{Institut f\"ur Organische Chemie und Biochemie, Technische
Universit\"at M\"unchen,\\ Lichtenbergstr.\ 4, D-85748 Garching,
Germany}

\maketitle

\vspace*{-1.5\baselineskip}
\begin{abstract}
We demonstrate a five-bit nuclear-magnetic-resonance quantum computer
that distinguishes among various functions on four bits, making use of
quantum parallelism.  Its construction draws on the recognition of the
sufficiency of linear coupling along a chain of nuclear spins, the
synthesis of a suitably coupled molecule, and the use of a multi-channel

spectrometer.
\end{abstract}

\section{Introduction}

While quantum computers of two bits have been implemented \cite{bitsa},
as have nuclear-magnetic-resonance (NMR) quantum computers of three
bits \cite{bitsc}, extending the number of bits has not proved easy.
We report the implementation of an NMR quantum computer having five
bits, involving the use of a linear coupling pattern
\cite{Brueschweiler}, synthesis of a molecule having five usable
spin-active nuclei with predominantly linear spin-spin coupling, and
the development of radio-frequency (r.f.) pulse sequences to act as
quantum logic gates for the molecule synthesized. Techniques to
suppress unwanted couplings between nuclear spins are described, as are
techniques to avoid perturbing some nuclear spins while manipulating
others.  Results are presented of a test of the five-bit computer on a
problem of Deutsch and Jozsa to distinguish one class of mathematical
function from another \cite{deutsch}.

\section{Definition of an {\bf \lowercase{n}}-bit NMR computer}

An $n$-bit quantum computer is called on to do three things: 1) accept
an instruction to prepare a starting state and prepare that state; 2)
accept instructions for and implement quantum gates (from which more
general unitary transformations of the state can be composed); and 3)
measure the state and yield an outcome.  The connection to computation
with classical computers depends on the recognition, due to
Bennett\cite{bennett2}, that all classical computations can be made
reversible.  Any terminating reversible computation is a permutation of
the inputs, which is unitary, and thus belongs to the class of
transformation performable on a quantum computer.  (For issues of
possibly nonterminating programs, see \cite{myers}.)\looseness=-1

In theory, a variant of the quantum computer is the expectation-value
quantum computer (EVQC), which in place of an outcome of a measurement
yields the expectation value \cite{gradientPPS1,gradientPPS2}.  NMR
quantum computing was born of the recognition that an EVQC can be
approximated by use of an NMR spectrometer containing a liquid sample,
the molecules of which have $n$ atoms with a nuclear spin of 1/2 (and
possibly other atoms, either spinless or having spins not used)
\cite{gradientPPS1,gradientPPS2,logicalPPS}.  Because tumbling of the
molecules decouples each molecule from all the others, the sample can be

described by a density matrix for the nuclear spins of the atoms of a
single molecule \cite{Ernst}, with only the spin-degrees of freedom,
corresponding to the desired Hilbert space of dimension $2^n$.  NMR
spectrometers sense only the traceless part of the density matrix, so
in place of matter in a pure state, an NMR computer can use a liquid
sample described by a density matrix proportional to a sum of a pure
state and any multiple of the unit matrix.  Such a density matrix,
called a {\it pseudopure} state \cite{gradientPPS1}, plays a role in
the 5-bit quantum computer.

Acting as an $n$-bit EVQC, a suitable NMR spectrometer allows the
preparation of a pseudopure starting state, the programming and
execution of r.f.\ pulse sequences that implement quantum gates, and the

determination of expectation values visible in NMR spectra.  To perform
the unitary operations required of a quantum computer, a sufficient set
of quantum gates consists of all single-spin operations and all
controlled-not gates that act on one nuclear spin under the control of
another nuclear spin.  Single-spin gates are implemented by selective
r.f.\ pulses.  Controlled-not gates between nuclei having spin-spin
coupling will be described, along with techniques to avoid unwanted
influences on other spins.  A key feature of the present design of the
NMR quantum computer is the reliance on a chain of linear coupling and
the use of swap gates to implement a controlled-not in which a spin $j$
controls spin $k$, where $j$ and $k$ have no direct spin-spin coupling
\cite{Brueschweiler}.  This allows use in NMR quantum computers of a
molecule having a simpler coupling pattern, and eases the problem of
unwanted influences on spins.

\section{Design of Test}

The proof of the pudding is in the eating: the 5-bit NMR computer to be
described was tested on the Deutsch-Jozsa problem for functions of 4
bits\cite{deutsch}, in the form described in \cite{cleve}, modified for
efficiency with NMR as described by Jones and Mosca \cite{bitsb}. (A
recent simplification \cite{collins}, unused here, would permit working
with functions of 5 bits.)  The problem is to decide whether a function
program selected from a set of possible programs computes one kind of
function or another.  Specifically, the problem is to distinguish
programs for balanced functions from programs for constant functions,
where the functions are from $\{0,1\}^4$ to $\{0,1\}$. (A function is
constant if its value is independent of its argument, and is called
balanced if the value for half the arguments is $1$ while the value is
$0$ for the other half.)  The test actually made was to distinguish
between programs for one constant and one balanced function, defined as
follows:
\begin{equation}
f_{0}(\vec{x}) \stackrel{\rm def}{=} 0\end{equation}
and
\begin{equation}
f_{b}(\vec{x}) \stackrel{\rm def}{=} x_{1}\oplus x_{2}\oplus
x_{3}\oplus x_{4}\label{eq:fb}
\end{equation}
for all $\vec{x}$, where $\vec{x} \stackrel{\rm def}{=}
(x_{1},x_{2},x_{3},x_{4})$, and ``$\oplus$'' is addition modulo 2.
Also, several controlled-not (CNOT) gates were tested, along with a
variety of 1-bit operators.  The balanced function chosen, $f_{b}$, has
the nice property of being implementable also in classical reversible
gates with no work bits.

Used to solve this problem, a quantum computer is a resource used both
to specify the function under test and to determine what it is.  In
order to separate these two uses, one can view the quantum computer as
used alternately by a {\it specifier} of the function and a {\it
decision maker}, two players of a game in which: (A) the decision maker
prepares the starting state; (B) the specifier runs the function
program;\footnote{These two moves must be iterated for a classical
computer, but not in the quantum solution of the Deutsch-Jozsa problem,
giving the quantum computer a large advantage over the classical
computer.} and (C) the decision maker makes a
measurement independent of the function program, and interprets the
result to decide the function class.

On an NMR quantum computer, (A) the decision maker starts a play by
using r.f.\ pulses and magnetic-field gradients (independent of the
function to be specified) to put the liquid sample in the pseudopure
state having a density matrix with a traceless part proportional to
\begin{equation} \rho_{i} \stackrel{\rm def}{=}16 |00001\rangle\langle
00001|-\frac{1}{2} {\bf 1} = 16
I_1^{\alpha}I_2^{\alpha}I_3^{\alpha}I_4^{\alpha}I_5^{\beta}-
\frac{1}{2}{\bf 1}\label{eq:rhoi}\end{equation} in terms of the
polarization operators $I_k^{\alpha} = (\frac{1}{2}{\bf 1}+I_{kz})$
and $I_k^{\beta} = (\frac{1}{2}{\bf 1}-I_{kz})$ usual to
NMR \cite{Ernst,Brueschweiler}.  Then the decision maker applies a
unitary transform $U_{90}$ by use of a hard
$90^{\circ}$ $y$-pulse which for this particular state has the same
effect
as the Hadamard transform on each spin \cite{bitsb}.
\begin{eqnarray} \rho_{i} \stackrel{U_{90}}{\rightarrow} \rho_0
&=& U_{90} \rho_i U_{90}^{\dag}\nonumber\\
& = &16 \left(\frac{1}{2}{\bf
1}+I_{1x}\right)\left(\frac{1}{2}{\bf 1}
+I_{2x}\right)\left(\frac{1}{2}{\bf
1}+I_{3x}\right)\nonumber\\
&&\mbox{}\times \left(\frac{1}{2}{\bf
1}+I_{4x}\right)\left(\frac{1}{2}{\bf
1}-I_{5x}\right) -\frac{1}{2}{\bf
1}.\label{eq:rho0}\end{eqnarray}

(B) The specifier chooses a function $f$ from one of the set of
functions undergoing test, here $f_{0}$ or $f_{b}$, and runs the
quantum version of a program to compute $f$; this program is a sequence
of gates, each a unitary transformation implemented by an r.f.\ pulse
sequence.  The total program implements a unitary transformation $U(f)$,

defined by its action on basis vectors $|\vec{x},\ x_5\rangle$:
\begin{equation} U(f)|\vec{x},\ x_5\rangle
= |\vec{x},\ x_5\oplus f(\vec{x})\rangle. \label{eq:ufunc}
\end{equation} The transform $U(f)$ produces the density matrix with
traceless part proportional to $\rho_f$:
\begin{equation}\rho_0 \stackrel{U(f)}{\rightarrow}
\rho_f.\end{equation}

(C) The decision maker reads out the NMR spectrum which depends
on $\rho_f$.  The spectrum differs according to whether $f$ is
balanced or constant, and thus tells the decision maker the function
class, with only one function evaluation, a large saving over
classical computation, which could require 9 evaluations for functions
of four bits.

In theory, for the case $f = f_0$, $U(f)$ is specified to be $U(f_0)$,
which by Eqs.\ (1) and (5) turns out to be the identity matrix, so one
should have $\rho_f = \rho_0$.  The spectrometer detects only the
terms of the righthand side of Eq.\ (\ref{eq:rho0}) that are linear in
$I_{x}$, so for a spectrometer adjusted to give an upward peak for
$I_{x}$, the resulting spectrum is in theory $\frac{|\,|\,|\,|\,
} {\;\;\;\;\;\;\;|}$, which has, from left to right, positive peaks for
spins 1 to 4 and a negative peak for spin~5.

For the balanced function, $f = f_{b}$ (Eq.\ (\ref{eq:fb})), $U(f_b)$ is

defined by $U(f_b)|\vec{x},\ x_5\rangle = |\vec{x},\ x_{1}\oplus
x_{2}\oplus x_{3}\oplus x_{4}\oplus x_5\rangle$.  A unitary operator
that is simpler to implement, that has the same effect on the fifth
(value) bit, and that allows the distinction between constant and
balanced functions is $\tilde{U}(f_b)$ defined by
\begin{eqnarray}\tilde{U}(f_b)|\vec{x},\ x_5\rangle &=& |x_{1},\ x_{1}
\oplus x_{2},\ x_{1} \oplus x_{2}\oplus x_{3},\ x_{1}\oplus
x_{2}\nonumber\\
&&\mbox{}\oplus
x_{3}\oplus x_{4},  x_{1}\oplus x_{2}\oplus x_{3}\oplus x_{4}\oplus
x_5\rangle,\label{eq:Vb}\end{eqnarray}
which we implemented by
sequential application of the gates (CNOT)$_{12}$, (CNOT)$_{23}$,
(CNOT)$_{34}$, and (CNOT)$_{45}$.  The spectrum calculated for the
density matrix $\tilde{\rho}_b$ obtained by transforming $\rho_0$ with
$\tilde{U}(f_b)$ is $\frac{|\,|\,|\,\;\; }{\;\;\;\;\;|\,|}$ ({\it vide
infra}).

\section{Realization and Test of a 5-bit NMR Computer}

A 5-bit NMR quantum computer requires a molecule having 5 spin-active
nuclei, with long relaxation times.  Large separation of resonance
frequencies of the nuclei allows rapid selective control of the spins.
For frequency separation, it is desirable to use different atomic
species for different spins, which requires a multi-channel NMR
spectrometer.  Our NMR experiments were performed
using a BRUKER AVANCE 400 spectrometer with five independent r.f.\
channels and a QXI probe (H,C-F,N). The lock coil was also used for
deuterium decoupling utilizing a lock switch.  A linear path of
spin-spin couplings is sufficient for all computations
\cite{Brueschweiler}.  Given the availability of a 5-channel
spectrometer, we chose as the ``hardware'' of our NMR quantum computing
experiments the molecule
BOC-($^{13}$C$_2$-$^{15}$N-$^{2}$D$_2^\alpha$-glycine)-fluoride which
contains an isolated coupling network consisting of five nuclei, each
having spin 1/2: the amide $^{1}$H, the $^{15}$N, the aliphatic
$^{13}$C$^\alpha$, the carbonyl $^{13}$C$^\prime$, and the $^{19}$F
nuclear spin (see Fig.~1). For simplicity, we will refer to these
spins (and the corresponding bits) as 1, 2, 3, 4 and 5, respectively.
All spins are heteronuclear, except for C$^\alpha$ and C$^\prime$
which however have a relatively large chemical shift difference.  The
five-spin system is well isolated from the protons of the BOC
protecting group which are separated by more than four chemical
bonds. In addition, the deuterium spins (D) which are attached to
C$^\alpha$ can be fully decoupled from the spins of interest using
standard heteronuclear decoupling techniques \cite{dec1,dec2,dec3}.  The

substance was synthesized starting from commercially available $^{13}$C
and $^{15}$N labeled glycine (see Appendix~A) and was dissolved in
deuterated dimethyl-sulfoxide (DMSO-D$_6$). NMR experiments were
performed at a magnetic field of about 9.4 Tesla
and a sample temperature of 27$^\circ$~C.  The experimentally
determined $T_2$ relaxation times for spins 1--5 were 250 ms, 490 ms,
450 ms, 590 ms, and 260 ms, respectively.  Resonance frequencies
$\nu_k$ and scalar coupling constants $J_{kl}$ are summarized in
Table~I. Except for the
$J_{23}$ coupling constant of 13.5 Hz, the spin chain is connected by
one-bond coupling constants $J_{k\{k+1\}}$ larger than 60 Hz.  In the
multiple rotating frame (see Ref.~\onlinecite{Ernst} and Appendix~B)
the precession frequency of each individual spin is 0, which
considerably simplifies implementation, because only coupling terms
need to be  considered (and manipulated).

\begin{table}
\caption{Resonance frequencies $\nu_{k}$, chemical shifts
$\delta_{k}$, one-bond coupling constants $J_{k\{k+1\}}$, and non-zero
two-bond coupling constants $J_{k\{k+2\}}$ of the used five-spin
system. No resolved three- or four-bond coupling constants were
observed.}
\vspace{8pt}
\vbox{%\columnwidth=3.5in
\begin{tabular}{ll}
$\nu_1= 400,133,001.6$ Hz & ($\delta_{1}= \hphantom{00}7.51$
ppm)\\[2pt]
$\nu_2= \hphantom{0}40,547,895.3$ Hz & ($\delta_{2}
=\hphantom{0}75.54$ ppm)\\[2pt]
$\nu_3= 100,616,858.0$ Hz & ($\delta_{3}= \hphantom{0}41.05$
ppm)\\[2pt]
$\nu_4= 100,629,089.1$ Hz & ($\delta_{4}=162.61$ ppm)\\[2pt]
$\nu_5= 376,510,545.5$ Hz & ($\delta_{5}= \hphantom{0}31.92$
ppm)\\[.5em]
$J_{12}= \hphantom{0}94.1$ Hz & $J_{23}= \hphantom{0}13.5$ Hz\\[2pt]
$J_{34}= \hphantom{0}65.2$ Hz & $J_{45}= 366.0$ Hz\\[0.5em]
$J_{13}= \hphantom{00}2.7$ Hz & $J_{35}= \hphantom{0}67.7$ Hz\\
\end{tabular}
}
\vskip1.5\baselineskip
\caption{List of initial Cartesian product operator terms of
$\rho_i$ (Eqs.~(3) and (\protect\ref{eq:terms})) that give rise to
detectable signals for at least one of the functions $f_0$ or $f_b$  in
the implemented version \protect\cite{cleve} of the Deutsch-Jozsa
algorithm (see also Appendices B and C). The propagators $U(f_0)={\bf
1}$ and $\tilde{U}(f_b)$ (Eq.\ (7)) transform $\rho_0$ (Eq.\ (4)) to
$\rho_0$ and $\tilde{\rho}_b$, respectively (Eq.\ (6)). Only the
underlined terms of $\rho_0$ and $\tilde{\rho}_b$ which contain single
transverse spin operators correspond to single quantum coherences that
are detectable in an NMR experiment.}

\vspace{8pt}
\vbox{%\columnwidth=3.5in
\begin{tabular}{ccc}
$\rho_i$&$\rho_0$&$\tilde{\rho}_b$\\[.7ex]
\hline
\noalign{\vskip3pt}
$I_{1z}$&$\underline{I_{1x}}$&$16 I_{1x} I_{2x} I_{3x} I_{4x}
I_{5x}$\\[.5ex]
$2 I_{1z} I_{2z}$&$2I_{1x} I_{2x}$&$\underline{I_{1x}}$\\[.5ex]
$I_{2z}$  & $\underline{I_{2x}}$ & $8  I_{2x} I_{3x} I_{4x}
I_{5x}$\\[.5ex]
$2 I_{2z} I_{3z}$& $2 I_{2x} I_{3x}$ & $\underline{I_{2x}}$\\[.5ex]
$I_{3z}$ & $\underline{I_{3x}}$  & $4  I_{3x} I_{4x} I_{5x}$\\[.5ex]
$2 I_{3z} I_{4z}$ & $2 I_{3x} I_{4x}$ & $\underline{I_{3x}}$\\[.5ex]
$I_{4z}$ & $\underline{I_{4x}}$ & $2  I_{4x} I_{5x}$\\[.5ex]
$-2 I_{4z} I_{5z}$&$-2 I_{4x} I_{5x}$&$\underline{-I_{4x}}$\\[.5ex]
$-I_{5z}$  & $\underline{-I_{5x}}$ & $\underline{- I_{5x}}$\\[.5ex]
\end{tabular}
}
\end{table}

\vskip-1.5\baselineskip
The experimental implementation of the propagator $U(f_0)$
corresponding to $f_0$ is trivial because by Eq.\ (\ref{eq:ufunc}) the
propagator is the unit operator, implemented by doing nothing.  In
contrast, the construction of the pulse sequence to implement the
series of CNOT-gates that define the unitary transformation
$\tilde{U}(f_b)$ of Eq.\ (\ref{eq:Vb}) for the balanced function $f_b$
(Eq.\ (\ref{eq:fb})) requires attention.  The goal is to create robust
pulse sequence elements that minimize the effects of experimental
imperfections. The pulse sequence elements shown in Fig.~2 A-D were
designed specifically for the coupling topology of our 5-spin system
to implement the unitary operators corresponding to \ ({CNOT})$_{12}$,
({CNOT})$_{23}$, ({CNOT})$_{34}$, and ({CNOT})$_{45}$.\footnote{During
each pulse sequence shown in Fig.\ 2 only the coupling $J_{kl}$ is
active which is required in order to implement (${\rm CNOT})_{kl}$.
The effects of other non-zero couplings (see Table I) are effectively
eliminated, except for $J_{13}$ in the sequence implementing
({CNOT})$_{45}$ (Fig.\ 2 D).  Although it would be straightforward to
remove also this coupling, this would require additional pulses which
can be avoided because in our spin system the coupling $J_{13}=2.7$ has
a negligible effect during the relatively short duration
$\Delta_{45}=1/(2 J_{45})=1.39$ ms of this gate.}
During these CNOT gates that act on two directly coupled spins $k$ and
$l$, only the couplings $J_{kl}$ are active, while the effect of all
other couplings in the spin system are refocused by cyclic pulse
sequences \cite{Ernst,Waugh,Quant}.  Figure~3 shows schematically the
pulse sequence actually used for the propagator $\tilde{U}(f_b)$ for
the balanced function $f_b$; this sequence benefited from applying
simple rules for pulse cancellation (see Appendix B.2).

The NMR implementation of the Deutsch-Jozsa algorithm starts with the
preparation of the pseudopure state $\rho_{i}$ of
Eq.\ (\ref{eq:rhoi}). The preparation of such a pseudopure state by a
single pulse sequence requires a non-unitary transformation of the
thermal equilibrium density operator \cite{gradientPPS1,bounds}.  This
can be achieved using spatial averaging
\cite{gradientPPS1,Brueschweiler} or temporal averaging
\cite{gradientPPS1,temporalPPS}.  In the basis formed by Cartesian
product operators \cite{Ernst}, $\rho_{i}$ can be expressed as a linear
combination of 31 terms that only consist of $z$ spin operators:
\begin{eqnarray}
\rho_{i}&=& \sum_{n=1}^5 s_n \, I_{nz} +\sum_{m<n} s_n \, 2 I_{mz}
I_{nz}  + \sum_{l<m<n} s_n\,  4 I_{lz} I_{mz} I_{nz}
\nonumber \\
&&\mbox{}+ \sum_{k<l<m<n} s_n\, 8 I_{kz} I_{lz} I_{mz}
I_{nz} - 16 I_{1z} I_{2z} I_{3z} I_{4z} I_{5z},\nonumber\\
\label{eq:terms}
\end{eqnarray}
where $s_n=-1$ if $n=5$ and $s_n=1$ otherwise. It is straightforward to
create each of these terms from the thermal equilibrium density
operator, using standard building blocks of high-resolution
NMR \cite{INEPT}.  In principle, temporal averaging could be realized
by repeating steps (A)--(C) of the game for all 31 terms in Eq.\
(\ref{eq:terms}) and by summing up the resulting spectra.  However,
because currently available NMR spectrometers require a distinct
experiment to detect each spin species ($^{1}$H, $^{15}$N, $^{13}$C
and $^{19}$F) (see Appendix B.3), a total of 124 NMR experiments
would be required for each function $f$ in order to include all terms
in the temporal averaging. A detailed analysis shows that of the 31
terms that constitute the pseudopure state $\rho_{i}$, only the five
linear terms $I_{kz}$ and the four bilinear terms $2I_{kz}
I_{\{k+1\}z}$ are transformed into detectable operators by the
propagator $U_{90}$ (to create $\rho_0$) followed by the propagators
$U(f_0)$ or $\tilde{U}(f_b)$, as the case may be (see Table~II and
Appendix~C).  As pointed out previously \cite{bitsc}, preparing just the

linear terms $I_{kz}$ suffices in some cases of the Deutsch-Jozsa
problem to distinguish constant from balanced functions, because in
these cases a balanced function gives a vanishing signal for at least
one of the input spins.  However, in the presence of experimental
imperfections, it is desirable to identify a balanced function based on
the sign reversal of the signal of at least one of the input spins,
rather than by the lack of a signal.  For the special case of the
balanced function $f_b$ that was chosen for this demonstration
experiment, this can be achieved by including also the bilinear terms
$2I_{kz} I_{\{k+1\}z}$ as starting operators (see Table~II).

Samples described by these linear and bilinear terms of $\rho_i$ were
prepared (see Appendix B.4) to demonstrate experimental control of the
five-spin system and to execute cases of the Deutsch-Jozsa algorithm.
For each function ($f_0$ and $f_b$) the following three sets of
experiments were performed (see experimental spectra in Fig.~4). Set 1
(first row of curves from the bottom in Fig.~4): preparation of the
linear terms
$I_{kz}$ (with algebraic signs as specified in Eq.~(8) and Table~II),
application of $U_{90}$ and $U(f)$, and detection of spin $k$ for
$k=1,\ \dots,\ 5$; set 2 (second row in Fig.~4): preparation of the
bilinear terms $2 I_{kz} I_{\{k+1\}z}$ (with algebraic signs as
specified in Eq.~(8) and Table~II), application of $U_{90}$ and $U(f)$,
and detection of spin $k$ for $k=1,\ \dots,\ 4$; and set 3 (third row in

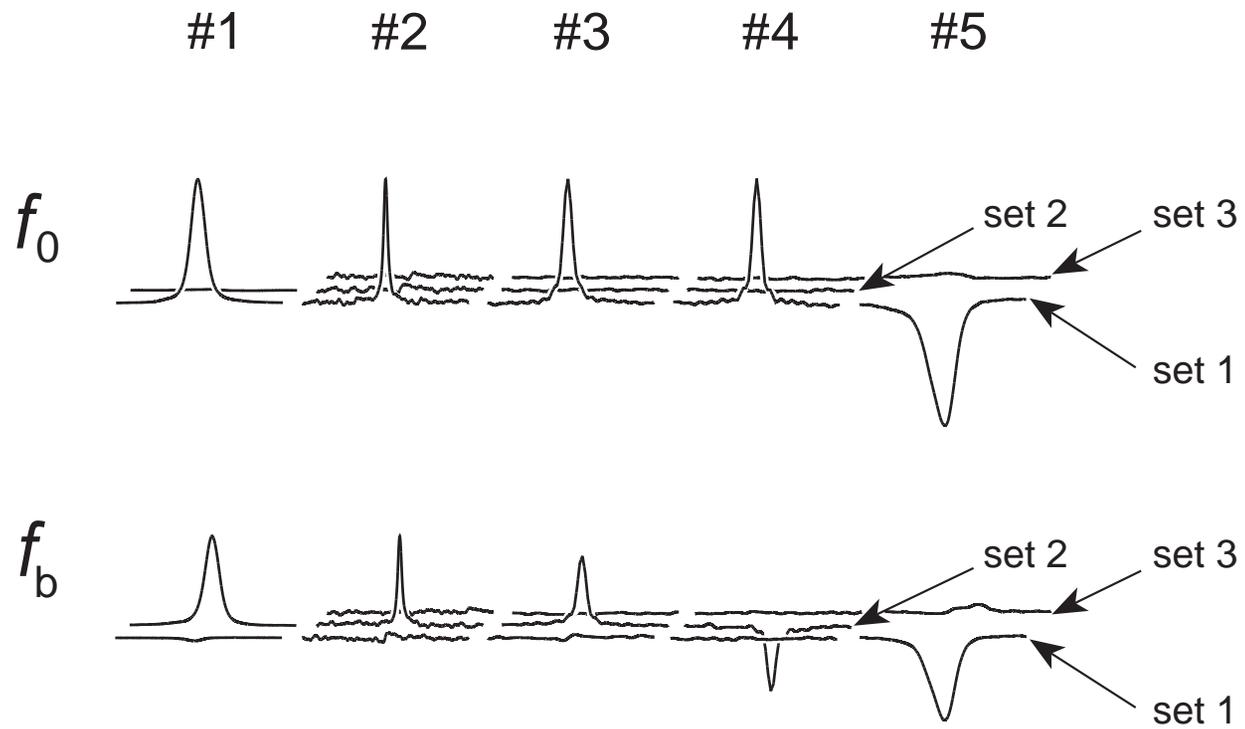
Fig.~4): preparation of the bilinear terms $2 I_{\{k-1\}z} I_{kz}$ (with

algebraic signs as specified in Eq.~(8) and Table~II), application of
$U_{90}$ and $U(f)$, and detection of spin $k$ for $k=2,\ \dots,\ 5$.

The observed spectra shown in Fig.~4 correspond closely to the
theoretical predictions (see Table~II). For the constant function
$f_0$, only the experiments of set~1 yield detectable signals.  For the
balanced function $f_b$, the experiments of set~1 only yield a
detectable signal for spin 5, whereas for spins 1--4 detectable signals
are only obtained in the experiments of set~2. As expected, only
spurious signals are detected for the experiments of set~3. The
amplitude of these spurious signals is typically on the order of 4\%
compared to the full signals.  As expected, all the signals of spins
1--4 are positive for the constant function whereas the signal of spin
4 is inverted by the propagator $\tilde{U}(f_b)$ corresponding to the
balanced function.  For $f_b$ the signal amplitudes reach only between
55\% and 70\% of the amplitudes found for $f_0$. This signal loss can
be attributed mainly to relaxation and experimental imperfections
during the sequence that implements $\tilde{U}(f_b)$ (Fig.~3), which
has an overall duration of 51.4 ms.

Through combined synthetic, analytic, and spectroscopic work, a
five-bit NMR quantum computer was built and shown to implement
superposition, quantum interference, and designed unitary
transformations. Although obstacles had to be overcome, none were
fundamental, and quantum computers with more than five bits will be
built.  Lots of interesting questions have been raised for future work
pertaining to the constraints and opportunities for linking molecular
architecture, spectrometer design, and algorithms for NMR quantum
computing.

\acknowledgments

S.J.G. acknowledges support by the Fonds der Chemischen Industrie and
the DFG. R.M. is supported by a stipend of the Fonds der Chemischen
Industrie and the Bundesministerium f\"ur Bildung und Forschung
(BMBF).  We thank C. Griesinger, M. Grundl, R. Kerssebaum, B. Luy, R.
Mayr-Stein, M. Kettner, M. Reggelin, H. Schwalbe, and A. T\"uchelmann
for valuable discussions and technical assistance.  A.F.F. thanks G.
Wagner (Harvard Medical School) for support and encouragement and
acknowledges support from National Science Foundation.  J.M.M. thanks
T.~T. Wu (Harvard University) for many critical insights.

%\newpage
\appendix

\def\thesubsection{\Alph{section}.\arabic{subsection}}

\section{Synthesis of Molecule}

We purchased 250 mg of $^{13}$C$_2$-$^{15}$N-glycine from Martek
Biosciences Corporation, 6480 Dobbin Road, Columbia, Maryland 21045.
The labeled glycine was fully deuterated by treatment with a solution
of NaOD in D$_2$O at 140$^\circ$C.  The product was dissolved in water
for reprotonation while retaining the deuterium atoms in
alpha-position.  The resulting
$^{13}$C$_2$-$^{15}$N-$^2$D$_2^\alpha$-glycine was protected in a
standard reaction with di-{\it tert}.-butyl-dicarbonate (BOC-anhydride)
as reagent (O. Keller, W.~E. Keller, G. van Look and G. Wersin,
Org.\ Synth.\ {\bf 63}, 160 (1985)).  Finally the carboxylic acid was
converted by cyanuric fluoride into the desired acyl fluoride:
BOC-($^{13}$C$_2$-$^{15}$N-$^{2}$D$_2^\alpha$-glycine)-fluoride (L.~A.
Carpino, E.~M.~E. Mansour and D. Sadat-Aalaee, J. Org.\ Chem.\
{\bf 56}, 2611 (1991)).  The substance dissolved in DMSO-D$_6$ at room
temperature shows NMR spectra that weaken with a half-life of about a
week, indicative of reactions not yet determined. The solution was
stable during storage at a temperature of $-30^\circ$~C.

\section{NMR pulse sequences}

For the preparation of the elements of a pseudopure state and the
implementation of quantum gates, robust r.f.\ pulse sequences are
desirable.  Pulse-sequence parameters with negligible experimental
errors are the durations of r.f.\ pulses and of delays. In addition,
the phases of r.f.\ pulses and of the receiver can be controlled with
negligible errors. The most important experimental imperfections are
r.f.\ amplitude errors that result from miscalibrations and from the
r.f.\ field inhomogeneity created by the r.f.\ coils.  In addition to
the use of compensating schemes, such as super cycles and composite
pulses \cite{Ernst}, experimental imperfections can be reduced by
designing pulse sequence elements with a minimum number of r.f.\
pulses.  For example, pulses to refocus frequency offset terms in
homonuclear spin systems with different chemical shifts can be
eliminated by implementing the experiments in the multiple-rotating
frame in which the precession frequency of each individual spin is 0
(see section B.1). More generally, pulses can often be eliminated or
replaced by phase adjustments with negligible errors (see section B.2).
For the available spectrometer, the experimental pulse parameters are
summarized in section B.3. The preparation of the elements of the
pseudopure state $\rho_i$ is discussed in section B.4.

\subsection{Implementation of experiments in the multiple rotating
frame}

For heteronuclear spins with resonance frequencies $\nu_k$ and $\nu_l$
in the laboratory frame, the spins are irradiated on-resonance and the
observed signals are demodulated by the determined resonance
frequencies.  If only a single r.f.\ channel is available for several
homonuclear spins, on-resonance irradiation of several homonuclear
spins can be achieved using phase-modulation of the r.f.\ pulses.  The
reference phase of each pulse applied to spin $k$ must be adjusted
such that it matches the desired phase in the corresponding rotating
frame ({\it vide infra}). In addition, the phases of the detected
signals need to be corrected for the relative phases that have been
acquired by the respective rotating frames during the course of the
experiment. In our case with the two homonuclear spins C$^{\alpha}$
(spin \#3) and C$^\prime$ (spin \#4), the transmitter frequency of the
carbon r.f.\ channel was set to the C$^{\alpha}$ resonance
frequency. In order to simplify the combination of different quantum
gates, the durations of the pulse sequences for each gate were chosen
to be integer multiples of $\Delta=1/\vert \nu_3-\nu_4\vert=81.75\
\mu$s.  Hence, the rotating frames are aligned at the end of each
gate.

\subsection{Simplifying pulse sequences}

Some quantum gates, such as (CNOT)$_{kl}$, require $z$ rotations of
individual spins which can be implemented using composite r.f.\
pulses \cite{compz}.  However, these pulses can be avoided if
$z$ rotations (by angle $\varphi$) are implemented by a corresponding
negative rotation
of the respective rotating frame of reference. In practice, this
results in an additional phase shift (by angle $- \varphi$) of all
following r.f.\ pulses that are applied to this spin and of the
receiver phase of this spin.  Furthermore, $180_\vartheta^\circ$
pulses (with arbitrary phase $\vartheta$) are required in some cases
to refocus the evolution due to $J$ couplings. In order to undo the
rotation caused by these pulses, additional $180_\vartheta^\circ$ or
$180_{-\vartheta}^\circ$ pulses are often needed at the beginning or
at the end of these quantum gates. An appropriate choice of the
position and phase $\vartheta$ of these pulses often makes it possible
to cancel two pulses from adjacent gates (e.g.,  $180^\circ_x$ and
$180^\circ_{-x}$) or to absorb a 180$^\circ$ pulse into the phase of
an adjacent 90$^\circ$ pulse (e.g., a $180^\circ_{x}$ pulse preceded or
followed by a $90^\circ_{-x}$ pulse is equivalent to a single
$90^\circ_{x}$ pulse).

Even if r.f.\ pulses cannot be completely eliminated, the accumulation
of small flip angle errors can be avoided by a proper choice of pulse
phases which is common practice in the design of modern NMR multiple
pulse sequences \cite{Ernst}.  For example,
the so-called MLEV-4 expansion \cite{dec1,dec2,dec3}
$180^\circ_{x}$$180^\circ_{-x}$$180^\circ_{-x}$$180^\circ_{x}$ (used
here, e.g., for spin 5 decoupling during spin 3- and spin
4-selective $90^\circ$ pulses) is preferable to
$180^\circ_{x}$$180^\circ_{-x}$$180^\circ_{x}$$180^\circ_{-x}$ or to
$180^\circ_{x}$$180^\circ_{x}$$180^\circ_{x}$$180^\circ_{x}$.

\subsection{Experimental pulse parameters}

Due to their large frequency separation, selective pulses for spins 1
($^{1}$H), 2 ($^{15}$N) and 5 ($^{19}$F) could be implemented by
simple square pulses. The durations of 90$^\circ$ pulses were 8.85
$\mu$s, 41 $\mu$s and 11.75 $\mu$s, respectively.  For spins 3
($^{13}$C$^\alpha$) and 4 ($^{13}$C$^\prime$) the following shaped
pulses with minimal durations and optimal selectivity were chosen
based on numerical simulations and experimental
optimizations: 90$^\circ$ pulses were implemented as e-SNOB
pulses \cite{esnob}, not for the usual 270$^\circ$, but for a $90^\circ$

rotation with a duration of 224
$\mu$s;
selective 180$^\circ$ pulses were
implemented as Gaussian pulses \cite{gauss} with a duration of 250
$\mu$s and a truncation level of 20\%.  The application of these
shaped e-SNOB and Gaussian pulses on C$^\alpha$ has a nonresonant
effect \cite{McCoy} on
C$^\prime$ which corresponds to experimentally determined $z$ rotations
of
$\varphi_e=-4^\circ$ and
$\varphi_g=-18^\circ$, respectively. Conversely, a shaped e-SNOB pulse
applied to C$^\prime$ leads to a $z$ rotation of $-\varphi_e $ for
C$^\alpha$. In all experiments these phase shifts were taken into
account by adjusting the phases of the following pulse and the
receiver phases (see Fig.~3).
(Note that the phases of the two selective Gaussian 180$^\circ$ pulses
applied to spin 3 in the period
$\tau_{45}$ is not corrected because their absolute phases are
arbitrary, c.f.\ Appendix~B.2.)  During the spin 3- or spin
4-selective 180$^\circ$ pulses, the evolution due to the strong
$J_{35}$ and $J_{45}$ couplings is automatically refocused. As this is
not the case for spin 3- or spin 4-selective 90$^\circ$ pulses, spin 5
was actively decoupled during these pulses (see Fig.~3).

As commercial high-resolution NMR spectrometers are commonly not
equipped with multiple receivers, it was not possible to
simultaneously detect the signals of different spin species. Moreover,
the application of any given pulse sequence required four different
pulse programs because the routing of the r.f.\ channels (for the
creation of $^{1}$H, $^{15}$N, $^{13}$C, $^{19}$F and $^{2}$D pulses)
depends on the detected spin species ($^{1}$H, $^{15}$N, $^{13}$C or
$^{19}$F).  Due to this technical limitation, each spin species had to
be detected in a separate experiment for every term of the initial
density operator $\rho_i$.  However, this made it possible to use
standard heteronuclear decoupling techniques to simplify the
detected signals and to significantly increase the signal-to-noise
ratio of the experiments.

During spin 1 detection, spins 2 and 3 were decoupled with an
r.f.\ amplitude $\nu_{rf}=\gamma B_{rf}/(2 \pi)$ of $0.6$ kHz and $0.4$
kHz, respectively.  During spin 2 detection, spins 1 and 3 were
decoupled with an r.f.\ amplitude of $2.3$ kHz and $0.4$ kHz,
respectively.  During spin 3 detection, spins 1, 2 and 5 were
decoupled with an r.f.\ amplitude of $2.3$ kHz, $0.6$ kHz and $2.0$
kHz, respectively, and during spin 4 detection, spin 5 was decoupled
with an r.f.\ amplitude of $2.0$ kHz.  In all these cases, the WALTZ-16
decoupling sequence \cite{dec1,dec2,dec3} was used.  In principle, also
the $J_{34}$ coupling could be effectively eliminated during detection
of spin 3 or spin 4 using time-shared decoupling. However, this was not
possible with our experimental setup because more than five separate
r.f.\ channels would have been required.  From the resulting doublets
(with splitting $J_{34}$) an apparent singlet was created by merging
the two doublet components \cite{sattler}.  During spin 5 detection,
spins 3 and 4 were simultaneously decoupled using a double-selective
G3-MLEV sequence \cite{dec1,dec2,dec3} with an r.f.\ amplitude of 6
kHz.  In addition, during all experiments deuterium decoupling was
applied using a WALTZ-16 sequence with $\nu_{rf}=0.5$ kHz.  In order to
approximate a constant sample temperature of about 27$^\circ$~C in
spite of the additional sample heating effected by the decoupling
sequences,  16 dummy scans were used prior to signal acquisition of
spins 1 and 5, whereas 4 dummy scans were used prior to signal
acquisition of spins 2, 3, and 4. Nevertheless, the linewidths of the
experimental signals shown in Fig.~4 were slightly increased by
residual sample heating effects and imperfections of the decoupling
sequences.

\subsection{Pulse sequences for the preparation of the terms of
$\rho_{i}$ and $\rho_{0}$}

In order to improve the sensitivity of the experiments and to filter
out signals from impurities in the sample, individual Cartesian product
operator terms of $\rho_{i}$ were created using sequential INEPT
transfer steps \cite{INEPT} starting from
$^1$H magnetization, corresponding to the operator $I_{1z}$. The term
$I_{1z}$ was prepared from the thermal equilibrium density operator by
applying spin 2, 3, 4, and 5 selective $90^\circ$ pulses followed by a
pulsed field gradient of the static magnetic field. An X-filter element
\cite{x-filt} was used to select $^{1}$H spins that are coupled to
$^{15}$N. For the preparation of other terms of $\rho_{i}$ (starting
from $I_{1z}$), the phase $\phi_a$ of the first 90$^\circ$ pulse
applied to spin 1 was subject to a two-step phase cycle. In addition,
the phase  $\phi_b$ of the 90$^\circ$ pulses for the implementation of
$U_0$ (see Eq.~(4)) was also subject to an independent phase cycle.
Overall, this resulted in a four-step phase cycle with the pulse phases
$\phi_a=\{0^\circ,180^\circ, 0^\circ, 180^\circ\}$,
$\phi_b=\{90^\circ, 90^\circ, 270^\circ, 270^\circ\}$ and the relative
receiver phases $\phi_{\rm rec}=\{0^\circ,180^\circ, 0^\circ,
180^\circ\}$.

\section{Density operator terms}

For all 31 Cartesian product operator terms in $\rho_i$
(Eq.\ (\ref{eq:terms})), the corresponding terms in $\rho_0$ and
$\tilde{\rho}_b$ are summarized in Table III.  The transformation $
\rho_0 \stackrel{\tilde{U}(f_b)}{\longrightarrow} \tilde{\rho}_b $
of the unitary operator corresponding to the balanced function $f_b$
is composed of four consecutive unitary transformations corresponding
to $({\rm CNOT})_{kl}$ quantum gates:
\begin{displaymath}
\rho_0
\ \stackrel{({\rm CNOT})_{12}}{\longrightarrow} \ \rho^\prime
\ \stackrel{({\rm CNOT})_{23}}{\longrightarrow} \ \rho^{\prime \prime}
\ \stackrel{({\rm CNOT})_{34}}{\longrightarrow} \ \rho^{\prime \prime
\prime}
\ \stackrel{({\rm CNOT})_{45}}{\longrightarrow} \ \tilde{\rho}_b.
\end{displaymath}

The transformations of the individual $({\rm CNOT})_{kl}$ gates can be
derived using the following rules \cite{bitsc}:
\begin{eqnarray*}
I_{kx}
& \stackrel{({\rm CNOT})_{kl}}{\longrightarrow}& 2 I_{kx}I_{lx},\\
2 I_{kx}I_{lx}& \stackrel{({\rm
CNOT})_{kl}}{\longrightarrow}& I_{kx},\\
I_{lx}& \stackrel{({\rm CNOT})_{kl}}{\longrightarrow}&
I_{lx}.
\end{eqnarray*}

The terms of the intermediate operators $\rho^\prime$,
$\rho^{\prime\prime}$ and $\rho^{\prime\prime\prime}$
are also given in Table III for completeness.

\end{document}